\documentclass{ws-procs975x65}

\usepackage{graphicx}

\newcommand{\beq}{\begin{equation}} 
\newcommand{\eeq}{\end{equation}} 

\newcommand{\nn}{\nonumber}

\begin{document}

\title{SELF-FORCE AND RADIAL FALL: NEW INTEGRATION METHOD FOR THE ZERILLI WAVE EQUATION}
\author{SOFIANE AOUDIA} 

\address{Max Planck Institut f\"ur Gravitationphysik, A. Einstein\\
Am M\"uhlenberg 1, 14476 Golm, Deutschland\\
E-mail: aoudia@aei.mpg.de}

\author{ALESSANDRO D.A.M. SPALLICCI}

\address{Observatoire des Sciences de l'Univers en R\'egion Centre, Universit\'e d'Orl\'eans \\
Campus CNRS, LPC2E, 3A Avenue de la Recherche Scientifique, 45071 Orl\'eans, France\\
E-mail: spallicci@cnrs-orleans.fr}

\begin{abstract}
We establish the jump conditions for the wavefunction and its derivatives through the formal solutions of the wave equation. These conditions respond to the requirement of continuity of the perturbations at the position of the particle and they are given for any mode at first order. Using these jump conditions, we then propose a new method for computing the radiated waveform without direct integration of the source term. We consider this approach potentially applicable to generic orbits.   
\end{abstract}


\bodymatter

\section{The jump conditions}

The complexity in assessing the continuity of the perturbations at the position of the particle in the Regge-Wheeler gauge has led, among other motivations, to work in the Lorenz gauge, at the price of loosing the availability of the wave equation. Nevertheless, it has been indicated by two different heuristic arguments \cite{lo00, lona09} that the even metric perturbations for radial fall should belong to the $C^0$ continuity class at the position of the particle, in the Regge-Wheeler gauge\footnote{The discontinuities of the wave function and its derivatives were also addressed elsewhere \cite{sola06}.}. 
Herein, we require that the perturbations are $C^0$ by identifying  the conditions that the wavefunction and its derivatives have to satisfy for allowing the perturbations to belong to such category. Our analysis is based on the solutions of the Zerilli equation, not on the equation themselves. 
The inverse relations for the perturbation functions $K$, $H_2$, $H_1$ are given by (having suppressed the $l$ index and being 
$m=0$):
\beq
K=\frac{6M^2+3M\lambda r+\lambda (\lambda +1)r^2}
{r^2(\lambda r+3M)}\Psi
+\left( 1-\frac{2M}r\right) \,\Psi_{,r}
-\frac{
\kappa \ U^0(r-2M)^2}{(\lambda +1)(\lambda r+3M)r}\delta 
\nn
\eeq

\[ 
H_2\!=\!-\!\frac{9M^3\!+\!9\lambda M^2r\!+\!3\lambda ^2Mr^2\!+\!\lambda ^2(\lambda \!+\!1)r^3}
{r^2(\lambda r\!+\!3M)^2}\,\Psi   
\!+\!
\frac{3M^2\!-\!\lambda Mr\!+\!\lambda r^2}{r(\lambda r\!+\! 3M)}\Psi_{,r}
\!+\! (r\!-\!2M)\Psi_{,rr} 
\]
\beq
+ \frac{\kappa U^0(r - 2M)(\lambda ^2r^2+2\lambda Mr-3Mr+3M^2)}{r (\lambda +1)(\lambda r+3M)^2}\delta -
\frac{\kappa U^0(r-2M)^2}{
(\lambda +1)(\lambda r+3M)}\delta'  
\nn
\eeq

\beq
H_1=\frac{\lambda r^2-3M\lambda r-3M^2}{\left( r-2M\right) (\lambda r+3M)}{\Psi_{,t}}+r\Psi_{,tr} 
- \frac{\kappa \ U^0\stackrel{.}{z}_u(\lambda r+M)}{(\lambda +1)(\lambda r+3M)}\delta 
+
\frac{\kappa \ U^0\stackrel{.}{z}_u r(r-2M)}{(\lambda +1)(\lambda r+3M)}\delta ' 
\nn
\eeq
where $ \lambda \!= \!1/2(l\! -\! 1)(l\! + \!2) $, $\kappa\! = \!4m\sqrt{(2l\!+\!1)\pi}$, $\delta \! = \! \delta\left[r\!-\!z_u(t)\right ]$ and $\delta'\! =\! \delta'\left[r\!-\!z_u(t)\right ]$; 
$U^0 \!= \!E/(1\! - \!2M/z_u)$ is the time component of the 4-velocity, $z_u$ the coordinate time dependent position, $E$ the energy of the particle. 
Since the wavefunction $\Psi$ belongs to the $C^{-1}$ continuity class, it and its derivatives can be written as, 
 
\beq 
\Psi(t,r)=\Psi^+(t,r)~\Theta_1+\Psi^-(t,r)~\Theta_2
~~~~~~~~~~~~~~~~
\Psi_{,r} = \Psi^+_{,r}\Theta_1 + \Psi^-_{,r} \Theta_2 + \left(\Psi^+ -\Psi^-\right) \delta
\nn
\eeq
\beq
\Psi_{,rr} = \Psi^+_{,rr}\Theta_1 + \Psi^-_{,rr} \Theta_2 + 
\left( \Psi^+_{,r} - \Psi^-_{,r} \right) \delta + 
\left(\Psi^+ - \Psi^-\right)\mid_{r = z_u} \delta'  
\nn
 \eeq
 

\beq
\Psi_{,t} = \Psi^+_{,t}\Theta_1 + \Psi^-_{,t} \Theta_2 - \left(\Psi^+ -\Psi^-\right) \dot{z}_u \delta
\nn
 \eeq


\beq
\Psi_{,tr} = 
 \Psi^+_{,tr}\Theta_1 + \Psi^-_{,tr} \Theta_2 + \left(\Psi^+_{,t} -\Psi^-_{,t}\right) \delta
  - \left(\Psi^+ -\Psi^-\right)\mid_{r = z_u} \dot{z}_u \delta'
\nn
 \eeq
where $\Theta_1 = \Theta\left[r-z_u(t)\right ]$, and $\Theta_2 = \Theta\left[z_u(t) - r \right ]$ are two Heaviside step distributions. For the second derivatives, the property of the Dirac delta distribution, at the position of the particle: $f(r)
\delta'[r-z_u(t)]=f(z_u(t)) \delta'[r-z_u(t)] - f'(z_u(t)) \delta[r-z_u(t)]$, has been used. 
The discontinuities of $\Psi$ and its derivatives must be such that they cancel when combined in $K$, $H_2$ and $H_1$ at the position of the particle. 
After replacing $\Psi$ and its derivatives in the perturbations, continuity requires that the coefficients of $\Theta_1$ must be equal to those 
of $\Theta_2$, while 
the coefficients of $\delta$ and $\delta '$ must vanish separately. Finally, the jump conditions for $\Psi$ and its derivatives are found (the jump conditions provided by 
$K$, $H_2$ and $H_1$ are equivalent):

\beq
\Psi^+ - \Psi^- =
\frac{\kappa E z_u}{(\lambda +1) (3 M+\lambda 
   z_u)}
~~~~~~~~~~~~~~
\Psi^+_{,r} - \Psi^-_{,r} =
\frac{\kappa E \left[6 M^2+3 M \lambda  z_u+\lambda 
   (\lambda +1) z_u^2\right]}{(\lambda +1) (2
   M-z_u) (3 M+\lambda  z_u)^2}
\nn
\eeq

\beq
\Psi^+_{,rr} - \Psi^-_{,rr}\! = \!
-\frac{\kappa E \left[3 M^3 (5 \lambda -3)+6 M^2 \lambda (\lambda -3)
    z_u+3 M \lambda ^2
(\lambda -1)    z_u^2-2 \lambda ^2 (\lambda +1)
   z_u^3\right]}{(\lambda +1) (2M-z_u)^2 (3
   M+\lambda  z_u)^3}
\nn
\eeq

\beq
\Psi^+_{,t} - \Psi^-_{,t}\! = \!
-\frac{\kappa E z_u \dot{z}_{u}}{(2 M - z_u) (3
   M+\lambda z_u)}
\nn
~~~~~~~~~~~~~
 \Psi^+_{,tr} - \Psi^-_{,tr} \!= \!
\frac{\kappa E \left(3 M^2+3 M \lambda 
   z_u - \lambda  z_u^2\right)\dot{z}_{u}}{(2M - z_u)^2 (3 M+\lambda  z_u)^2}
\nn
\eeq

\section{The new integration method}

The key idea of this method is to use the above jump conditions (rewritten in terms of $r_*$ and $t$) rather than integrating the source term over the cells which are crossed by the world line of the particle. Instead, for the cells never crossed by the world line, the integration method retains the classic procedure \cite{lopr97b, mapo02}. We define $\alpha=\Psi(t+h,r_*)$, $\beta=\Psi(t,r_*-h)$, $\gamma=\Psi(t-h,r_*)$ and $\delta=\Psi(t,r_*+h)$ as the four vertices of the diamond of Fig.~1 centered on $\sigma=\psi(t,r_*)$.
For the case of Fig.~1, the world line crosses the line joining $\beta$ and $\delta$ at the point $a$ and the line joining $\gamma$ and $\alpha$ at the point $b$. We get a set of 8 equations relating   
$\Psi_a^+$, $\Psi_a^-$, $\Psi_b^+$, $\Psi_b^-$, $\Psi^+_{,r_*}$, $\Psi^-_{,r_*}$, 
$\Psi^+_{,t}$, $\Psi^-_{t}$, 
$\alpha$, $\beta$, $\gamma$, $\delta$, $\sigma$, $\epsilon_a$, $\epsilon_b$, $h$, from which we deduce the value of $\alpha$: 

\beq
\alpha = \beta + \delta - \gamma  + (\Psi^+ - \Psi^-)_b - (\Psi^+ - \Psi^-)_a + \epsilon_b~(\Psi^+_{,t} - \Psi^-_{,t})_b -\epsilon_a~(\Psi^+_{,r_*} - \Psi^-_{,r_*})_a
\nn
\eeq

There are other three different ways that the particle may cross the cell and similar relations may be drawn. The method is applicable to generic orbits. 

\begin{figure}[t]
\begin{center}
\includegraphics[width=1.5in]{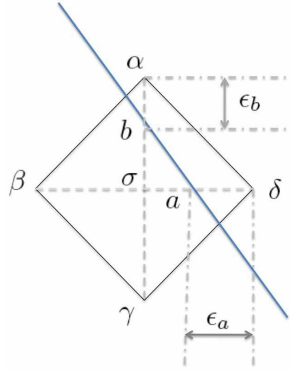}
\end{center}
\caption{One of the possible four cases that a cell is crossed by the particle.}
\label{aba:fig1}
\end{figure}

\section*{Acknowledgements}

Discussions with J. Mart\'in Garc\'ia (IAP, Paris) and M.-T. Jaekel (ENS, Paris) have clarified the jump conditions.  
The authors wish to acknowledge the FNAK (Fondation Nationale Alfred Kastler), the CJC (Conf\'ed\'eration des Jeunes Chercheurs) and all organisations which stand against discrimination of foreign researchers.

\end{document}